\documentstyle[twoside,fleqn,espcrc2]{article}

\newcommand{\re}[1]{(\ref{#1})} 

\newcommand{\ice}[1]{\relax}

\newcommand{\Break}{ \right. \nonumber \\ &{}& \left. }
\newcommand{\dsp}{\displaystyle}

\def\beq{\begin{equation}}
\def\eeq{\end{equation}}

\def\bea{\begin{eqnarray}}
\def\eea{\end{eqnarray}}

\def\ba{\begin{array}}
\def\ea{\end{array}}

\def\bq{\begin{quote}}
\def\eq{\end{quote}}

\def\nnb{\nonumber}

\newcommand{\msbar}{\overline{\mbox{MS}}} 
\newcommand{\RI}{\mbox{RI }} 
 
\newcommand{\MOM}{MOM}

\newcommand{\Si}{\Sigma}

\newcommand{\ep}{\epsilon}

\newcommand{\als}{\alpha_s}
\newcommand{\aspi}{\frac{\alpha_s}{\pi}}
\newcommand{\prd}{\partial}

\ice{
\newcommand{\ice}[1]{\relax}
\newcommand{\prd}{\partial}
\newcommand{\ep}{\epsilon}
\newcommand{\beq}{\begin{equation}}
\newcommand{\eeq}{\end{equation}}
\newcommand{\bea}{\begin{eqnarray}}
\newcommand{\eea}{\end{eqnarray}}

\newcommand{\re}[1]{(\ref{#1})}

\def\bbuildrel#1_#2^#3%
{\mathrel{\mathop{\kern 0pt#1}\limits_{#2}^{#3}}}

\newcommand{\gsim}{\;\rlap{\lower 3.5 pt \hbox{$\mathchar \sim$}} \raise 1pt
 \hbox {$>$}\;}
\newcommand{\lsim}{\;\rlap{\lower 3.5 pt \hbox{$\mathchar \sim$}} \raise 1pt
 \hbox {$<$}\;}

}

\catcode`\@=11
\def\slash{\mathpalette\make@slash}
\def\make@slash#1#2{\setbox\z@\hbox{$#1#2$}%
  \hbox to 0pt{\hss$#1/$\hss\kern-\wd0}\box0}
\catcode`\@=12 

\ice{

\newcommand{\ba}{\begin{array}}
\newcommand{\ea}{\end{array}}

}

\ice{

\newcommand{\msbar}{\overline{\mbox{MS}}}
  }
\ice{
\newcommand{\dsp}{\displaystyle}

}

\title{
\vskip-3cm{\baselineskip14pt
\centerline{\normalsize\hfill TTP99--39}
\centerline{\normalsize\hfill hep-ph/9909565}
\centerline{\normalsize\hfill September 1999}
}
\vskip.7cm
${\cal O}(\alpha_s^3)$
conversion relation  betweeen  $\msbar$ and Euclidean
quark masses\footnote{Invited talk given by K.G.Ch. at the QCD 99
Euroconference, Montpellier, France 7-13th July 1999} 
\author{K. G. Chetyrkin${}^{a,b}$ and A. R\'etey${}^b$
\\ { \vspace{.5cm}
{ ${}^{a}$
Institute for Nuclear Research, 
 Russian Academy of Sciences, \\
 60th October Anniversary Prospect 7a,
 Moscow 117312, Russia }
 \\ {\vspace{.5cm}
{${}^{b}$  Institut f\"ur Theoretische Teilchenphysik,
   Universit\"at Karlsruhe, 
   D-76128 Karlsruhe, Germany}
  }}}}

\begin{document}

\begin{abstract} 
We report on the analytical calculation of NNNLO (${\cal O}(\als^3)$)
conversion factor between the $\msbar$ quark mass and the one defined
in the so-called ``Regularization Invariant'' scheme. The
NNNLO contribution in the conversion factor turns out to be
relatively large and comparable to the known NNLO term.
\end{abstract}

\maketitle

\section{Introduction}\label{sec:intro}

\renewcommand{\thefootnote}{\fnsymbol{footnote}}

Quark masses\footnotetext[1]{Invited talk given by K.G.Ch. at the QCD 99
Euroconference, Montpellier, France 7-13th July 1999} 
 are fundamental parameters of the QCD Lagrangian.
Nevertheless, their relation to measurable physical quantities is not
direct: the masses depend on the  renormalization scheme and,
within a given one, on the renormalization scale $\mu$.

In the realm of pQCD the most often used definition is based on the
$\msbar$-scheme \cite{ms,MSbar} which leads to the so-called
short-distance $\msbar$ mass.  Such a definition is of great
convenience for dealing with mass-dependent inclusive physical
observables dominated by short distances (for a review see
\cite{CKKrep}).
Unfortunately, as their mass dependence is relatively weak the
predictions are usually difficult to use for getting a precise
information on quark masses.

To determine the absolute values of quark masses, one necessarily has
to rely on the methods which incorporate the features of
nonperturbative QCD. So far, the only two methods which are based on
QCD from the first principles are QCD sum rules and lattice QCD (for
recent discussions see e.g.
\cite{Prades:1997vn,Jamin:1997sa,Dominguez:1998zs,
Chetyrkin:1998ej,Prades:1998xa,Lubicz:1998kc,Kenway:1998ew,%
Gimenez:1998pu,Sharpe:1998hh}).  Rather accurate determinations of the
ratios of various quarks masses can be obtained within Chiral
Perturbation Theory \cite{leut1}.

Lattice QCD provides a direct way to determine a quark mass from first
principles.  Unlike QCD sum rules it does not require model
assumptions.  It is possible to carry out the systematic improvement
of lattice QCD so that all the discretization errors proportional
to the lattice spacing are eliminated (a comprehensible review is
given in \cite{Luscher:1998pe}). The resulting quark mass is the (short
distance) bare lattice quark mass. The matching of the lattice quark
masses to those defined in a continuum perturbative scheme requires
the calculation of the corresponding multiplicative renormalization
constants.  In the \RI scheme \cite{NMPmethod} the renormalization
conditions are applied to amputated Green functions in Landau gauge,
setting them equal to their tree-level values.  This allows
the non-perturbative calculation of the renormalization constants. An
alternative to the \RI approach is the Schroedinger functional scheme
(SF) which was used in \cite{Capitani:1997mw,Sint:1998iq}

Once the RI quark masses are determined from lattice calculations they
should be related to the $\msbar$ mass by a corresponding conversion
factor. By necessity this factor can be defined and, hence, computed
only {\em perturbatively}. The conversion factor is now known at
next-to-next-to-leading order (NNLO) from
Ref.~\cite{Franco:1998bm}. The NNLO contribution happens to be
numerically significant. This makes mandatory to know the NNNLO
$O(\alpha_s^3)$ term in the conversion factor.

In this work   we report  on  the calculation of this term. It
turns out that the size of the newly computed term is comparable to
the previous one at a renormalization scale of 2 GeV --- the typical
scale currently used in the lattice calculations of the light quark
masses. This means that perturbation theory can not be used for a
precise conversion of  the presently available \RI quark masses
to the $\msbar$ ones. A simple analysis shows that the convergency
gets   better if the scale is increased to, say, 3  GeV. 
Thus, once the lattice calculations produce  the \RI quark masses
at this scale our formulas will allow accurate conversion to
the $\msbar$ masses at  the same scale. 
\section{Scheme dependence of quark mass}
\label{sec:massfield}
We start by considering the bare quark propagator (for simplicity we
stick to the Landau gauge  and, thus, do not explicitly
display the gauge dependence)
\beq
\label{def:fermi}
F_0(q,\alpha_s^0,m_0)
 =  (m_0  - \slash{q} - \Si_0)^{-1}
{},
\eeq
with the quark mass operator $\Si_0$ being conveniently decomposed
into Lorentz invariant structures according to
\beq
\label{def:sigmas}
\Si_0 = \slash{q} \Si^0_V + m_0 \Si^0_S
\nnb {}.
\eeq
Here $m_0$ and $\psi_0$ is the bare quark mass and field respectively:
$
\dsp
a_s
\equiv  \frac{\alpha_s}{\pi}
= \frac{g^2}{4\pi^2}
\, 
$
and $g$ is the bare QCD gauge coupling.  To be precise we assume that
\re{def:fermi} is dimensionally regulated by going to non-integer
values of the space-time dimension $D=4-2\ep$
\cite{dim.reg-a,dim.reg-b}. The $\msbar$ renormalized counterpart of
the Green function \re{def:fermi} reads
\beq
\label{def:fermi-ren}
\ba{l}
\dsp
F(q,\alpha_s,m,\mu) 
=  (m  - \slash{q} - \Si)^{-1} 
\\
\dsp
=Z_2^{-1} F_0(q,\alpha_s^0,m_0)|_{{}_{
\scriptstyle
m_0=Z_m m,\, \alpha_s^0= \mu^\ep Z_\alpha \alpha_s 
                                     }
                                }
{},
\ea
\eeq
where the renormalized quark field 
$
\psi= Z_2^{-1/2} \psi_0
{}
$
and the {}'Hooft mass parameter $\mu$ is a scale at which the
renormalized quark mass is defined.  The renormalization constants
$Z_2,Z_\alpha$ and $Z_m$ are series of the generic form ( $? =
2,\alpha$ or $m$ )
\beq 
Z_? =
1 + \sum_{i>0} Z_?^{(i)} \frac{1}{\ep^i}, \, \, \, 
\, \,  Z_?^{(i)} = \sum_{ j \ge i } Z_?^{(i,j)}
\left( \aspi \right )^j
{}.
\label{def:Z-expansions}
\end{equation}
Now let us consider the quark propagator renormalized according to a
different subtraction procedure. Marking with a prime parameters of
the second scheme one can write
\beq
\label{def:fermi-ren-prime}
\ba{l}
\dsp
F(q,\alpha'_s,m',\mu) = 
 \frac{1}{m'  - \slash{q} - \Si'}
\\
=(Z'_2)^{-1} F_0(q,\alpha_s^0,m_0)|_{{}_{
\scriptstyle
m_0=Z'_m m, \,  \alpha_s^0= \mu^\ep Z_\alpha' \alpha_s 
                                     }
                                }
{},
\ea
\eeq
where without essential loss of generality we have set $\mu' = \mu$.
The finiteness of the renormalized fields and parameters in both
schemes implies that, within the framework of perturbation theory, the
relation between them can be uniquely described as follows
\beq
m = C_m \cdot m'
, \, \,\,
\psi = \sqrt{ C_2} \cdot \psi'  
\label{def:C2}
{},
\eeq
with the ``conversion functions'' being themselves {\em finite} series
in $\alpha_s'$, i.e.
\begin{equation}
C_?  \equiv        
1 + \sum_{i>0} C_?^{(i)} \left( \frac{\alpha_s'}{\pi}\right)^i 
{}
\label{def:C-expansions-ap}
\end{equation}
for $?=m$ or $2$.

Note that in general the coefficients $C_?^i$ may depend on the ratio
$m'/\mu$.  If such a dependence is absent then the corresponding
subtraction scheme is referred to as a ``mass independent'' one.  In
what follows we  limit ourselves to considering the latter
case. In addition, being only interested in conversion functions $C_2$
and $C_m$, we will assume that the function $C_\alpha$ has already
been determined and, thus, will deal with the following representation
of $C_2$ and $C_m$ in terms of the $\msbar$ coupling constant
$\alpha_s$ ( $? = 2,m$ )
\begin{equation}
C_?  \equiv        
1 + \sum_{i>0} C_?^{(i)} \left( \aspi\right)^i 
{}
\label{def:C-expansions-a}
\end{equation}

{}From eqs.~(\re{def:fermi}) it is easy to see that 
\bea
C_2 \cdot ( 1 + \Si_V ) = 1 + \Si'_V  
{},
\label{form:C2}
\\
C_2 \cdot C_m \cdot ( 1 - \Si_S ) = 1 - \Si'_S 
{}.
\label{form:Cm}
\eea
The renormalization conditions for the non-$\msbar$ scheme should then
be used to provide the necessary information about the right hand side
to calculate the conversion factors $C_m$ and $C_2$ given the $\msbar$
renormalized $\Si_V$ and $\Si_S$.

An example of a mass independent $\MOM$ scheme has recently been
suggested in \cite{NMPmethod} under the name of \RI (``Regularization
Invariant'') scheme.  The corresponding versions of the general conversions 
formulas read \cite{preparation} ( $\ell = log(-\frac{q^2}{\mu^2})$)
\bea
C^{RI}_2 = \left[ 
1 + \Si_V + \frac{1}{2}\cdot \frac{\partial\Si_V(\ell)}{\partial \ell} 
\right]^{-1}_{q^2=-\mu^2,\, m=0}
{},
\label{form:C2RI}
\\
C^{RI}_m = \left[ 
\frac{ 1 + \Si_V + \frac{1}{2} \cdot \frac{\partial\Si_V(\ell)}{\partial \ell}}
     { 1 - \Si_S } \right]_{q^2=-\mu^2\, m=0}
\label{form:CmRI}
{}.
\eea
\section{Three Loop Conversion functions}

We have analytically computed the functions  $\Sigma_V$ and
$\Sigma_S$  in the massless limit  to order $\alpha_s^3$.  
The calculation has
been done with intensive use of computer algebra programs.  In
particular, we have used {\tt QGRAF}~\cite{qgraf} for the generation
of diagrams, and {\tt MINCER}~\cite{mincer2} for their evaluation.

Our results for the conversion functions 
read (in the  Landau gauge and in terms of fractions
and Riemann's Zeta function and as functions of $n_f$):
\begin{eqnarray} 
C_2^{\RI} &=&   
1 + \left(\frac{\alpha_s}{4\pi}\right)^{2}
\left[
-\frac{517}{18} 
+12  \,\zeta_{3}
+\frac{5}{3}  \,n_f 
\right]
\nonumber\\
 &{+}& \left(\frac{\alpha_s}{4\pi}\right)^{3}
\left[
-\frac{1287283}{648} 
+\frac{14197}{12}  \,\zeta_{3}
\Break
+\frac{79}{4}  \,\zeta_{4}
-\frac{1165}{3}  \,\zeta_{5}
+\frac{18014}{81}  \,n_f 
\Break
\nonumber
-\frac{368}{9}  \,\zeta_{3} \,n_f 
-\frac{1102}{243}  \, n_f^{2}
\right]
{},
\label{C2RIqcd}
\end{eqnarray}

\begin{eqnarray}C_m^{\RI} &=& 
1  +   \frac{\alpha_s}{4\pi}
\left[ 
-\frac{16}{3}\right] 
\nonumber
\\
&+&  \left(\frac{\alpha_s}{4\pi}\right)^{2}
\left[
-\frac{1990}{9} 
+\frac{152}{3}  \,\zeta_{3}
+\frac{89}{9}  \,n_f 
\right]
\nonumber\\
 &{+}& \left(\frac{\alpha_s}{4\pi}\right)^{3}
\left[
-\frac{6663911}{648} 
+\frac{408007}{108}  \,\zeta_{3}
\Break
-\frac{2960}{9}  \,\zeta_{5}
+\frac{236650}{243}  \,n_f 
-\frac{4936}{27}  \,\zeta_{3} \,n_f 
\Break
\nonumber
+\frac{80}{3}  \,\zeta_{4} \,n_f 
-\frac{8918}{729}  \, n_f^{2}
-\frac{32}{27}  \,\zeta_{3} \, n_f^{2}
\right]
{}.
\label{CmRIqcd}
\end{eqnarray}

At a scale $\mu = 2$ GeV and $n_f=4$, the numerical contributions of
various orders are as follows (for simplicity we took
$\alpha_s(2 \, \mbox{GeV}) /\pi = .1$)

\beq
C_m^{RI} = 1.-0.1333-0.0754-0.0495
{}
\label{Cm}
\eeq
and
\beq
C_2^{RI} = 1.+0.- 0.00477 - 0.00508
\label{C2}
{}.
\eeq

One observes that the sizes of the NNLO and NNNLO contributions to
$C_m^{RI}$, at this scale amount to about 7.5\% and 5\%
respectively. This means that perturbation theory can not be used for
a precise conversion of the \RI quark masses to the $\msbar$ ones at
the renormalization scale $\mu = $ 2 GeV. The convergency can be
improved if one icreases $\mu$ till, say, 3 GeV. 
Indeed, with this choice of $\mu$ the standard three-loop evolution
gives $\alpha_s(3 \mbox{ GeV}) = 0.262$ and eqs. (\ref{Cm},\ref{C2})
transform to 

\beq
C_m^{RI} =  1. - 0.111  - 0.0526   - 0.0289 
{}
\label{Cm_3}
\eeq
and
\beq
C_2^{RI} = 1. - 0.00333  - 0.00296 
\label{C2_3}
{}.
\eeq

To conclude: the newly computed NNNLO corrections are numerically
significant and should be taken into account when transforming the
\RI quark masses to the $\msbar$ ones.

\section{Acknowledgements}
One of us (K.G.Ch.) is grateful to Damir Becirevic for illuminating
discussions and good advice. This work was supported by DFG under
Contract Ku 502/8-1 ({\it DFG-Forschergruppe ``Quantenfeldtheorie,
Computeralgebra und Monte-Carlo-Simulationen''}) and the {\it
Graduiertenkolleg ``Elementarteilchenphysik an Beschleunigern''}.

\def\app#1#2#3{{ Act.~Phys.~Pol.~}{ B #1} (#2) #3}
\def\apa#1#2#3{{ Act.~Phys.~Austr.~}{#1} (#2) #3}
\def\cmp#1#2#3{{ Comm.~Math.~Phys.~}{ #1} (#2) #3}
\def\cpc#1#2#3{{ Comp.~Phys.~Commun.~}{ #1} (#2) #3}
\def\epjc#1#2#3{{ Eur.\ Phys.\ J.\ }{ C #1} (#2) #3}
\def\fortp#1#2#3{{ Fortschr.~Phys.~}{#1} (#2) #3}
\def\ijmpa#1#2#3{{ Int.~J.~Mod.~Phys.~}{ A #1} (#2) #3}
\def\jcp#1#2#3{{ J.~Comp.~Phys.~}{ #1} (#2) #3}
\def\jetp#1#2#3{{ JETP~Lett.~}{ #1} (#2) #3}
\def\mpl#1#2#3{{ Mod.~Phys.~Lett.~}{ A #1} (#2) #3}
\def\nima#1#2#3{{ Nucl.~Inst.~Meth.~}{ A #1} (#2) #3}
\def\npb#1#2#3{{ Nucl.~Phys.~}{ B #1} (#2) #3}
\def\nca#1#2#3{{ Nuovo~Cim.~}{ #1A} (#2) #3}
\def\plb#1#2#3{{ Phys.~Lett.~}{ B #1} (#2) #3}
\def\prc#1#2#3{{ Phys.~Reports }{ #1} (#2) #3}
\def\prd#1#2#3{{ Phys.~Rev.~}{ D #1} (#2) #3}
\def\pR#1#2#3{{ Phys.~Rev.~}{ #1} (#2) #3}
\def\prl#1#2#3{{ Phys.~Rev.~Lett.~}{ #1} (#2) #3}
\def\pr#1#2#3{{ Phys.~Reports }{ #1} (#2) #3}
\def\ptp#1#2#3{{ Prog.~Theor.~Phys.~}{ #1} (#2) #3}
\def\sovnp#1#2#3{{ Sov.~J.~Nucl.~Phys.~}{ #1} (#2) #3}
\def\tmf#1#2#3{{ Teor.~Mat.~Fiz.~}{ #1} (#2) #3}
\def\tmph#1#2#3{{ Theor.~Math.~Phys.}{ #1} (#2) #3}
\def\yadfiz#1#2#3{{ Yad.~Fiz.~}{ #1} (#2) #3}
\def\zpc#1#2#3{{ Z.~Phys.~}{ C #1} (#2) #3}
\def\ibid#1#2#3{{ibid.~}{ #1} (#2) #3}

\end{document}